\documentclass[%
 reprint,
 amsmath,amssymb,
 aps,
]{revtex4-1}
\usepackage{graphicx}
\usepackage{dcolumn}
\usepackage{bm}
\usepackage{amsmath,scalerel}
\usepackage{subfigure}
\usepackage{subfigmat}
\usepackage{mathtools,tikz,caption}
\usepackage{soul}
\usepackage{amsmath}

\DeclareRobustCommand\sampleline[1]{%
  \tikz\draw[#1] (0,0) (0,\the\dimexpr\fontdimen22\textfont2\relax)
  -- (2em,\the\dimexpr\fontdimen22\textfont2\relax);%
}

\makeatletter
\newcommand*\bigcdot{\mathpalette\bigcdot@{.5}}
\newcommand*\bigcdot@[2]{\mathbin{\vcenter{\hbox{\scalebox{#2}{$\m@th#1\bullet$}}}}}
\makeatother

\begin{document}

\preprint{APS/123-QED}

\title{A Morphing Continuum Analysis of Energy \textcolor{black}{Transfer} in Compressible 
Turbulence}

\author{Mohamad Ibrahim Cheikh}
 \author{Louis B. Wonnell}
\author{James Chen}%
 \email{Corresponding Author, jmchen@ksu.edu}
\affiliation{%
Multiscale Computational Physics Lab\\
 Mechanical and Nuclear Engineering Department, Kansas State University, Manhattan, KS 66502, USA}%


\begin{abstract}
A shock-preserving finite volume solver with the generalized Lax-Friedrichs splitting flux for Morphing Continuum Theory (MCT) is presented and verified. The numerical MCT solver is showcased in a supersonic turbulent flow with Mach 2.93 over an $8^{\circ}$ compression ramp. The simulation results validated MCT with experiments as an alternative for modeling compressible turbulence. The required size of the smallest mesh cell for the MCT simulation is shown to be almost an order larger than that in a similar DNS study. The comparison shows MCT is a much more computationally friendly theory than the classical NS equations. The dynamics of energy cascade at the length-scale of individual eddies is illuminated through the subscale rotation introduced by MCT. In this regard, MCT provides a statistical averaging procedure for capturing energy transfer in compressible turbulence, not found in classical fluid theories. Analysis of the MCT results show the existence of a statistical coupling of the internal and translational kinetic energy fluctuations with the corresponding eddy rotational energy fluctuations, indicating a multiscale transfer of energy. In conclusion, MCT gives a new characterization of the energy cascade within compressible turbulence without the use of excessive computational resources.
\end{abstract}

\maketitle


\section{\label{sec:level1}Introduction}

Turbulence remains as one of the most relevant unsolved problems in physics 
today. The study of compressible turbulence, in particular, applies to many 
fields within physics and engineering, including supersonic aircraft design, 
inertial confinement fusion, and star formation within galaxies. Still, modeling 
and analyzing these complex flows is a constant challenge. 
\textcolor{black}{Kov{\'a}sznay addressed this challenge by decomposing the 
weak turbulent fluctuations about a uniform mean flow with spatially uniform 
mean thermodynamic properties into three modes of fluctuations; vortical, 
acoustic, and entropic modes \cite{kovasznay1953}. For first order modes,  the 
three types of disturbances are decoupled from each other at fluctuation 
amplitudes \cite{chu1958non}. For the second order modes, however, couplings 
arise between any two modes, and their interaction generates the other modes 
\cite{Monin1971Statistical}. Goldstein \cite{goldstein_1978} showed that for 
linear unsteady disturbances about an arbitrary potential flow, the 
fluctuations need only be decomposed into vortical and entropic modes. The 
vortical modes are found in divergence-free velocity fields with no pressure 
fluctuations, while the entropic mode arises from temperature spottiness 
\cite{kovasznay1953}. Kov{\'a}szany decomposition has been employed in linear 
rapid-distortion theory \cite{hunt1990rapid} in  homogeneous 
\cite{sagaut2008homogeneous} and inhomogeneous \cite{nazarenko1999} turbulence, 
and shock wave turbulent interactions \cite{andreopoulos1987some}. }

 The influence of the \textcolor{black}{vortical fluctuation modes or small scale } eddies on the dynamics of compressible turbulent flows is one of the most difficult aspects to simulate and visualize. The interactions between individual eddies and between subscale eddies and translational mean flow can provide insight into the energy cascade process \cite{leonard1975energy, frisch1978simple, moin1991dynamic, ertesvaag2000eddy, samtaney2001direct, aluie2012conservative, vaghefi2015local}. For the case of compressible turbulence, the interactions at the smallest length scales are fundamental to shock-turbulent boundary layer interactions (STBLI), when turbulence is amplified and eventually dissipated after passing through the shock wave \cite{dolling2001fifty, bookey2005new, wonnell2016morphing2, 
oliver2007assessment, pirozzoli2006direct, ducros:99jcp}.

  Kolmogorov's picture of a continuous transition of kinetic energy at the large scales to dissipation of heat at the molecular level still shapes the mainstream discussion of energy transfer within turbulence \cite{frisch1995turbulence}. A constant question, however, is the extent of this model's applicability to the smallest relevant length scales. Leonard employed filtering techniques to the incompressible Navier-Stokes equations to determine the contribution of subgrid-scale eddies to the energy cascade process \cite{leonard1975energy}. The nonlinear advection term was determined to be a primary factor in extracting energy from the mean flow, while the Reynolds stress component played a minor role. In the case of isothermal compressible turbulence, Aluie found evidence that Kolmogorov's picture may still be applicable when pressure dilatation effects decay sufficiently quickly \cite{aluie2012conservative}. A key part of Aluie's study was the observed statistical decoupling of kinetic and internal energy at smaller scales, allowing for local conservative cascade to the smaller eddies. Indeed, the energy transfer from the inertial length scale to the subscale eddies dramatically affects the dynamics at the smallest relevent length scales. The details of subscale motion then become important for
either molecular dissipation or inverse energy cascade.

From these studies, the specifics of the contributions of individual eddies are inferred from manipulations of the Navier-Stokes equations. Once a relevant smallest length scale is specified, the simulation or experiment cannot directly describe the dynamics of smaller eddies \cite{lee1997interaction}. DNS simulations can produce energy spectra for a wide range of length-scales \cite{kaneda2003energy, del2004scaling} but will inevitably incur higher computational costs if the details of individual eddies are needed. These limitations arise due to the assumption of the fluid as a continuum of infinitesimal points. Small-scale dynamics such as eddy rotation are inferred from the behavior of these points. Furthermore, the variables of the Navier-Stokes equations do not explicitly include terms that allow for the control of subscale motion. Velocity fields present useful data, but the interpretation of this subscale behavior is left to the researcher.

To extract details at the smallest scales, some researchers approached turbulence from a different starting assumption of the fluid. Eringen derived new equations for fluids containing an inner structure, where the components of the fluid possess a finite size and orientation \cite{eringen1966theory}. This new picture of the fluid, known as a morphing continuum, led to a mathematical formulation that incorporated a new term for the rotation of these inner structures. Since this formulation, the extent of the success of Morphing Continuum Theory (MCT) in reproducing turbulent profiles for compressible and incompressible turbulence has been well documented \cite{wonnell2016morphing2, chen2012micropolar, chen2012numerical, wonnell2017morphing, cheikh2017morphing, wonnell2017extension}. In particular, MCT was able to capture a post-shock inverse energy cascade through spectral analysis of the kinetic energies of translation and subscale rotation \cite{wonnell2016morphing2}. Still, a thorough application of MCT to the problem of energy cascade in compressible turbulence has yet to occur. 

This paper applies MCT to the problem of the contribution of subscale eddies to the energy cascade process for compressible turbulence. \textcolor{black}{Using MCT, the study is able to decompose the motion of the subscale eddies into translational and rotational motions, and investigates the energy transfer between kinetic and internal energy.} Supersonic freestream turbulence over a compression ramp is simulated and \textcolor{black}{used to analyze the energy transfer at the subscale level,}  using the governing equations and new relevant variables of MCT. Section \ref{sec:level2} gives the physical picture of the fluid through the lens of MCT and derives the relevant governing equations. Section \ref{sec:level3} describes the numerical scheme employed to discretize the equations of MCT and the algorithm to solve these equations. Section \ref{sec:level4} tests the order of accuracy of this numerical scheme on a simple Couette flow. Section \ref{sec:level5} describes the test case of the compression ramp, the results from the MCT simulation, and any relevant observations to the discussion of energy \textcolor{black}{transfer} for compressible turbulence. A further discussion and concluding thoughts are presented in section \ref{sec:level6}.

\section{Morphing Continuum Theory \label{sec:level2}}

Eringen's microcontinuum field theories \cite{eringen1966theory, 
eringen2001microcontinuum, eringen2001microcontinuum2,  eringen1972theory}, the 
starting physical picture for MCT, assume that the fluid is comprised of inner 
structures with arbitrary shapes and self-spinning rotation. The macroscopic and 
subscale motions of these inner structures are expressed by:
\begin{align}
x_{k}=x_{k}(X_K,t),\qquad k = 1,2,3 \\
X_{K} = X_K(x_{k},t), \qquad  K = 1,2,3 \\
\xi_{k}=\chi_{kK}(X_K,t)\Xi_K, \quad \Xi_K  = \bar{\chi}_{kK}\xi_{k}
\end{align}
where $X_K$ and $x_k$ refer to the initial and final positions of the inner 
structure, with coordinates denoted by $k$ or $K$, while $\xi_{k}$ and 
$\chi_{kK}$ represent the local rotation and deformation vectors of the inner 
structure. The inner structures have a total of nine degrees of 
freedom, making the mathematics extremely tedious.
MCT simplifies these inner structures by assuming subscale isotropy in 
deformation, thus removing the degrees of freedom related to the deformation 
$\chi_{kK}$ of the inner structures \cite{wonnell2016morphing2}. This 
simplification means that any deformation is presumed to be isotropic at the 
length-scale of the smallest inner structures. The resulting fluid element in 
MCT differs from the classical fluid theory by having, in addition to the 
translational motion, a local rotation characterized by the gyration, 
$\omega_{k}$. The angular momentum of these inner structures becomes $\rho j 
\omega_{k}$, where $j$ represents the inertia of the inner structure defined by 
Chen et al \cite{chen2012numerical,chen2011constitutive} to be,\begin{equation}
j= \frac{j_{mm}}{3}
\end{equation}
In addition to the isotropic deformation of the inner structures, 
Morphing Continuum Theory considers these inner structures to be small eddies in 
the turbulent flow. The theory assumes these eddies to have rigid spherical 
structures and constant material properties. Chen \cite{chen2012numerical} 
showed that the inertia of these spheres has the relation $j = \frac{2}{5}d^2$, 
where $d$ represents the sphere's diameter.  

The total velocity of these sub-grid eddies can now be written as ${v}^{total}_k 
= {v}_k + (d/2) \omega_k$, where $v_k$ is the eddy translational velocity and 
$\omega_k$ is the gyrational motion of an eddy. If the magnitude of the 
gyrational motion of the eddy is small compared to the translational motion, the 
gyration is mathematically equivalent to the perturbed velocity found in the 
Reynolds-Averaged Navier Stokes equations \cite{wonnell2016morphing2}.
\subsection{Governing Equations for Compressible Flow}
As discussed previously, the main variables that govern the motion of the subscale eddies in an MCT flow are the translational motion, $v_k$, and local gyration, $\omega_k$. From these variables, the new deformation rate of the MCT-based fluid becomes \cite{eringen1972theory}:
\begin{align}
a_{kl} &=v_{l,k} +\epsilon_{lkm}\omega_m \\
b_{kl} &=\omega_{k,l}
\end{align}
where $a_{kl}$ represents the classical deformation-rate tensor from the 
Navier-Stokes equations with an additional term representing the effect of the rotation of the inner structures. The $b_{kl}$ tensor is a new deformation tensor not found in the classical fluid theory, representing the strain experienced due to gradients in the gyration. \textcolor{black}{Decomposing the first deformation rate tensor into symmetric and skew-symmetric components yields,
\begin{align}
\textcolor{black}{a_{kl} = \underbrace{\frac{1}{2} (v_{l,k} + v_{k,l})}_{S_{kl}} + \underbrace{ \frac{1}{2} (v_{l,k} - v_{k,l} + 2\epsilon_{lkm}\omega_m)}_{\Omega_{kl}} }
\end{align}
where $S_{kl}$ represents the strain experienced by the deformation of the fluid 
element, similar to the classical Navier-Stokes theory, while $\Omega_{kl}$ 
represents the rigid body rotation. $\Omega_{kl}$ is similar to the classical 
spin rate tensor, but with an additional term that takes into account the effect 
of the gyration. Since $\Omega_{kl}$ is skew-symmetric, then the permutation 
tensor $\epsilon_{lkm}$ can be used to convert $\Omega_{kl}$ into a vector 
field. Chen \cite{chen2012numerical} referred to this field as the absolute 
rotational field, $\Omega_{k}$, characterized by the expression, 
\begin{align}
\Omega_{k} = \epsilon_{lkm} \Omega_{kl} = - \epsilon_{klm} v_{l,k} + 2 \omega_m
\end{align} 
For classical fluids, converting the spin rate tensor to a vector 
field yields a Galilean invariant vorticity. In MCT, $\Omega_{kl}$ yields a 
more general rotational field that includes the contribution of the gyration in 
addition to the vorticity field, making the absolute rotation field Galilean 
invariant and objective \cite{chen2012numerical}. Similar to the classical 
Navier-Stokes vorticity field, the absolute rotational field represents twice 
the rotation vector of the MCT fluid element.}

With the deformation-rate tensors above, the constitutive equations for the Cauchy stress tensor, moment stress tensor, and heat flux are derived to be: \cite{chen2012numerical}
\begin{align}
t_{kl} &= -p \delta_{kl} + \lambda \text{tr}(a_{mn})\delta_{kl} + (\mu + \kappa)a_{kl} + \mu a_{lk}\\
m_{kl} & = \alpha_T\epsilon_{klm} T_{,m} + \alpha 
\text{tr}(b_{mn})\delta_{kl} + \beta b_{kl} + \gamma b_{lk} \\
q_{k} & = -K T_{,k} + \alpha_{T}\epsilon_{klm}\omega_{m,l}
\end{align}
\noindent where $\rho$ is the fluid density; $p$ is the pressure; $\mu$ is the 
dynamic viscosity; $\lambda$ is the second coefficient of viscosity; $\kappa$ is 
the coupling coefficient between the linear and angular momenta; $\gamma$ is the 
subscale diffusion coefficient; $T_{,k}$ is the temperature gradient, and $K$ is 
the thermal conductivity; $\alpha_T$, $\alpha$ and $\beta$ are material 
constants \textcolor{black}{that are set to zero for this study}. Plugging these equations into the balance laws, one obtains the governing equation for a compressible flow:\\
\\
\noindent \textit{Conservation of Mass:}
\begin{equation}
\frac{D \rho}{D t} + \rho v_{m,m} = 0
\label{ConM}
\end{equation}
\textit{Conservation of Linear Momentum:}
\begin{equation}
\begin{split}
\rho \frac{D{v_m}}{Dt}= -p_{,m} + (\lambda + \mu) v_{n,nm} &+ (\mu + \kappa) v_{m,nn} \\
& + \kappa ( \epsilon_{mnk} {\omega}_{k,n} )
\end{split}
\label{LM}
\end{equation}
\textit{Conservation of Angular Momentum:}
\begin{equation}
\begin{split}
\rho j\frac{D \omega_m}{Dt}= (\alpha+\beta) \omega_{n,nm} &+ \gamma \omega_{m,nn} \\
&+ \kappa (\epsilon_{mnk} v_{k,n} - 2 \omega_m)
\label{ConAM}
\end{split}
\end{equation}
\textit{Conservation of Energy:}
\begin{align}
\begin{split}
\rho \frac{D E}{D t} &= -(p v _{m})_{,m} + ( \lambda v_{m,m} v_k )_{,k} + \left[ \kappa ( v_{l,k} v_{l}  \right. \\
& \left. + \epsilon_{kml}\omega_{m}v_{l}) + \mu ( v_{l,k} + v_{k,l} )v_{l} \right]_{,k}   \\ 
& + \left( \alpha \omega_{m,m} \omega_{k} + \beta \omega_{k,l} \omega_{l} + \gamma \omega_{l} \omega_{l,k} \right)_{,k} \\
& + \left(KT_{,k} \right)_{,k}
\end{split}
\label{ConE}
\end{align}
where $E = e + 1/2 (v_m v_m + j \omega_m \omega_m)$ is the total energy density 
of the fluid , and $e$ is the internal energy. $\alpha_T$ disappears after 
substitution into the balance laws, since $m_{kl,k}$ and $q_{k,k}$ will yield 
$\frac{\alpha_T}{T}e_{klm}T_{,mk} = 0$ and
$\frac{\alpha_T}{T}e_{klm}\omega_{m,lk} = 0$ .To close this system of equations 
the fluid is assumed to be an ideal gas, leading to the following 
relations:\begin{align}
e &= c_v T = c_v \frac{p}{\rho (c_p - c_v)} \\ 
\rho E &= \frac{p}{\frac{c_p}{c_v} - 1} + \frac{1}{2} \rho (v_m v_m + j \omega_m \omega_m) 
\end{align}
where $c_p$ is the specific heat at constant pressure and $c_v$ is the specific heat at constant volume. Finally the generalized Stokes' hypothesis for MCT is employed to relate the second coefficient of viscosity ($\lambda$) with dynamic viscosity ($\mu$) and coupling coefficient ($\kappa$) as \cite{chen2012numerical},
\begin{align}
3 \lambda + 2 \mu + \kappa = 0
\end{align}

\subsection{Nondimensional Form of the MCT Governing Equation}
To better understand the contribution of the individual eddies, the MCT governing equations be non-dimensionalized, where the dimensionless groups are defined based on the physical parameters of interest. Starting with the distance and motion variables, the length scales $x_m$, and the translation velocity $v_m$ will be parameterised with the  square-root of the subscale inertia $L = \sqrt{j}$, and the freestream velocity $U_\infty$ respectively. The temporal term $t$ will be dimensionalized with the time it takes the freestream velocity to cover the distance $L$, i.e. $L/U_\infty$. The gyration, $\omega_m$, meanwhile, will be dimensionalized with the inverse of temporal term. In summary the dimesionless variables are:
\begin{align}
\begin{split}
& \hat{x}_m = \frac{x_m}{L} \quad \hat{v}_m = \frac{v_m}{U_\infty} \\
& \hat{t} = \frac{t}{L/U_\infty} \quad \hat{w}_m = \frac{w_m}{U_\infty/L}
 \end{split}
\end{align}
The thermodynamic variables of the density, $\rho$, and pressure, $p$, will be dimensionalized according to the freestream density $\rho_\infty$ and dynamic pressure $\rho U_\infty^2$. Substituting the nondimensionalized variables into the governing equations yields a set of dimensionless groups that captures the physical behavior of each parameter. One parameter is the Reynolds number, which is defined as the ratio of the convection to the diffusion of linear momentum,
\begin{align}
Re = \frac{\rho_\infty U_\infty L}{\mu+ \kappa}
\end{align}
As for the energy equation two dimensionless numbers appear; the Prandtl number, which defines the ratio of momentum diffusivity to thermal diffusivity, and the Eckert number, which defines the relationship between a flow's kinetic energy and the boundary layer enthalpy difference,
\begin{align}
Pr = \frac{c_p (\mu + \kappa)}{K} \qquad Ec= \frac{U_\infty^2}{c_p T_\infty}
\end{align}
The previously defined parameters are typical dimensionless groups found in the classical fluid theory. The next dimensionless term that is specific to MCT, will be called $Er$ in honor of Eringen and is defined as the ratio of the inertial forces to the viscous forces arising from the gyration, 
\begin{align}
Er = \frac{\rho_\infty U_\infty L}{\kappa}
\end{align}
The other parameters found in MCT will also be non-dimensionlised with respect to the convection term
\begin{align}
C_\alpha = \frac{\rho_0 U_\infty L^3}{\alpha} \quad C_\beta = \frac{\rho_0 U_\infty L^3}{\beta} \quad & C_{\gamma} = \frac{\rho_0 U_\infty L^3}{\gamma} 
\end{align}
In this regard the governing equations in dimensionless form become:\\

\noindent \textit{Conservation of Mass:}
\begin{equation}
\frac{\partial \hat{\rho} }{\partial \hat{t}} +  \hat{\nabla} \bigcdot \left(\hat{\rho} \hat{\textbf{v}} \right) = 0
\label{DimCon}
\end{equation}

\noindent \textit{Conservation of Linear Momentum:}
\begin{align}
\begin{split}
\frac{\partial (\hat{\rho} \hat{\textbf{v}})}{\partial \hat{t}} & + \hat{\nabla} \bigcdot \left( \hat{\textbf{v}} (\hat{\rho} \hat{\textbf{v}}) \right) = - \hat{\nabla} \hat{p} + \frac{1}{Re}\left[ \frac{1}{3} \hat\nabla (\hat\nabla \bigcdot \hat{\textbf{v}}) \right. \\
& \left. + \hat\nabla^2 \hat{\textbf{v}}\right] + \frac{1}{Er}\left[\hat\nabla \times \boldsymbol{\hat\omega} -\frac{2}{3} \hat\nabla (\hat\nabla \bigcdot \hat{\textbf{v}})\right] 
\end{split}
\label{DimLin}
\end{align}

\noindent \textit{Conservation of Angular Momentum:}
\begin{align}
\begin{split}
\frac{\partial (\hat{\rho} \hat{\boldsymbol{\omega}})}{\partial \hat{t}} & + \hat{\nabla}\bigcdot \left( \rho \hat{\textbf{v}} \hat{\boldsymbol{\omega}} \right) = \left(\frac{1}{C_\alpha} + \frac{1}{C_\beta}\right) \hat{\nabla} (\hat{\nabla} \bigcdot \hat{\boldsymbol{\omega}} ) \\
& +  \frac{1}{C_\gamma} \hat{\nabla}^2 \hat{\boldsymbol{\omega}} + \frac{1}{Er}\left(\hat{\nabla} \times \hat{\textbf{v}} - 2 \hat{\boldsymbol{\omega}}\right)
\end{split}
\label{DimAng}
\end{align}

\noindent \textit{Conservation of Energy:}
\begin{align}
\begin{split}
\frac{\partial \hat{\rho} \hat{E}}{\partial \hat{t}}  &+\hat{\nabla}\bigcdot( \hat{\rho} \hat{E} \hat{\textbf{v}}) = -\hat{\nabla} \bigcdot (\hat{p}\hat{\textbf{v}})  + \frac{1}{Re} \hat{\nabla} \bigcdot \left[ [\hat{\nabla}\hat{\textbf{v}}]^T\bigcdot \hat{\textbf{v}} \right.
\\
& \left. + \hat{\textbf{v}} \bigcdot \hat{\nabla} \hat{\textbf{v}}  - \frac{2}{3} \hat{\textbf{v}}(\hat\nabla \bigcdot \hat{\textbf{v}})\right]  + \frac{1}{Er} \hat{\nabla} \bigcdot \left[ \frac{1}{3}\hat{\textbf{v}}(\hat\nabla \bigcdot \hat{\textbf{v}}) \right. 
\\
& \left.+  \hat{\boldsymbol{\omega}} \times \hat{\textbf{v}}  - [\hat{\nabla}\hat{\textbf{v}}]^T\bigcdot \hat{\textbf{v}}  \right] 
 + \frac{1}{C_\alpha} \hat{\nabla} \bigcdot ((\hat{\nabla} \bigcdot \hat{\boldsymbol\omega}) \hat{\boldsymbol\omega} )
  \\
&+ \frac{1}{C_\beta} \hat{\nabla} \bigcdot ((\hat{\nabla} \hat{\boldsymbol\omega})\bigcdot \hat{\boldsymbol \omega})  + \frac{1}{C_\gamma} \hat{\nabla}\bigcdot \left( \hat{\boldsymbol\omega} \bigcdot (\hat{\nabla} \hat{\boldsymbol\omega})^T\right)\\
&+\frac{1}{Re Ec Pr}\hat \nabla^2 \hat T
\end{split}
\label{DimEne}
\end{align}

\section{Numerical Scheme \label{sec:level3}}
The solver developed to implement the MCT compressible governing equations is be constructed in the framework of the finite volume discretization. One reason for choosing finite volume is due to its easy implementation, and its convergence to a stable solution for complex flows. The spatial domain implemented is divided into contiguous control volumes or cells, with the physical variables of velocity, gyration, pressure, density and temperature collocated (i.e. located at the cell center). 

\noindent The  transport equation for any conserved property can be written in following form,
\begin{equation}
 \underbrace{\frac{\partial \phi}{\partial t}}_\text{transient term} +  \underbrace{\nabla \bigcdot (\textbf{v} \phi)}_\text{convective term} = \underbrace{\nabla \bigcdot (\Gamma_\phi \nabla \phi )}_\text{diffusive term}  +\underbrace{S_\phi}_\text{source term}
\label{te}
\end{equation}
Here, $\phi$ refers to a transport variable, $\Gamma_\phi$ is the diffusivity or the diffusion coefficient, and $S_\phi$ is the source term. Letting $\phi = \rho$ yields the continuity equation, $\phi = \rho v_m$ gives the linear momentum equation, $\phi = j \rho \omega_m$ yields the angular momentum equation and $\phi = \rho E$ gives the energy equation. The finite volume method requires that the governing equations in their integral form be satisfied over the control volume. Applying spatial integration on equation \ref{te},
\begin{equation}
\int_{V_c} \frac{\partial \phi}{\partial t} dV + \int_{V_c}  \nabla \bigcdot (\textbf{v} \phi) dV = \int_{V_c} \nabla \bigcdot (\Gamma_\phi \nabla \phi) dV  \int_{V_c} S_\phi dV
\end{equation}
For the present solver, a simple forward Euler was implemented for the unsteady term,
\begin{equation}
\int_{V_c} \frac{\partial \phi}{\partial t} dV = \frac{\phi_c^{n+1} - \phi_c^{n}}{\bigtriangleup t} V_c
\end{equation}
where $V_c$ represents the cell volume, the subscript $c$ refers to the cell center, and superscript $n$ refers to the current time step. Implementing forward Euler on the conservation of mass, linear momentum, angular momentum, and energy equations yields:
\begin{align}
& \int_{V_c}\frac{\partial \hat{\rho} }{\partial \hat{t}}dV \approx \frac{\hat{\rho}_c^{n+1} - \hat{\rho}_c^{n}}{\bigtriangleup \hat{t}} V_c 
\\
& \int_{V_c}\frac{\partial (\hat{\rho} \hat{\textbf{v}})}{\partial \hat{t}} dV \approx \frac{(\hat{\rho} \hat{\textbf{v}})_c^{n+1} - (\hat{\rho} \hat{\textbf{v}})_c^{n}}{\bigtriangleup \hat{t}} V_c 
\\
& \int_{V_c}\frac{\partial (\hat{\rho} \hat{\boldsymbol{\omega}})}{\partial \hat{t}} dV \approx \frac{(\hat{\rho} \hat{\boldsymbol{\omega}})_c^{n+1} - (\hat{\rho} \hat{\boldsymbol{\omega}})_c^{n}}{\bigtriangleup \hat{t}} V_c 
\\
&\int_{V_c} \frac{\partial (\hat{\rho} \hat{E}) }{\partial \hat{t}} dV \approx \frac{(\hat{\rho} \hat{E})_c^{n+1} - (\hat{\rho} \hat{E})_c^{n}}{\bigtriangleup \hat{t}} V_c 
\end{align} 
This scheme is first order in time, but can be modified to a higher-order Runge-Kutta time integration scheme. 

Critical care is considered for the numerical scheme implemented on the convection terms in MCT, which are $\hat{\nabla} \bigcdot \left(\hat{\rho} \hat{\textbf{v}} \right)$,$\hat{\nabla}\bigcdot \left( \rho \hat{\textbf{v}} \hat{\textbf{v}} \right)$, $\hat{\nabla}\bigcdot \left( \rho \hat{\textbf{v}} \hat{\boldsymbol{\omega}} \right)$, and $\hat{\nabla}\bigcdot( \hat{\rho} \hat{E} \hat{\textbf{v}})$. The numerical scheme adopted for the convection terms should be able to capture the shock wave and discontinuities, while avoiding oscillations. Replacing the volume integral by a surface integral through the use of the divergence theorem, the convection terms can be approximated as,
\begin{equation}
\int_{V_c} \nabla \bigcdot (\textbf{v} \phi) dV = \oint_S (\textbf{v} \phi) \bigcdot d\textbf{S} \approx \sum_f \textbf{v}_f \phi_f \bigcdot \textbf{S}_f
\end{equation}
where $\sum_f$ denotes the summation over the faces of the control volume, 
$\textbf{v}_f \bigcdot \textbf{S}_f$ is the volumetric flux, $\textbf{S}_f$ is 
the face normal vector, and $\phi_f$ represents the face value of the transport 
variable. Notable methods found in the literature are able to effectively 
produce accurate non-oscillatory solutions for $\phi_f$. These methods are:  
piecewise parabolic method (PPM) \cite{colella1984piecewise}; essentially 
non-oscillatory (ENO) \cite{shu1988efficient,  harten1987uniformly}; weighted 
ENO (WENO) \cite{liu1994weighted}; and the Runge-Kutta discontinuous Galerkin 
(RKDG) method \cite{cockburn1998runge}. All of these methods involve Riemann 
solvers, characteristic decomposition and Jacobian evaluation, making them 
troublesome to implement. The scheme implemented in this study is a second-order 
semi-discrete, non-staggered scheme, introduced by Kurganov, Noelle and Petrova 
(KNP) \cite{kurganov2001semidiscrete} as a second-order 
generalized Lax-Friedrichs scheme. The interpolation procedure of the transport 
variable $\phi$ from the cell center, $\phi_c$, to the face center, $\phi_f$, 
implemented in this scheme is split into two directions corresponding to the 
outward or inward direction of the face normal,\begin{equation}
\begin{split}
\sum_f \textbf{v}_f \phi_f & \textbf{S}_f  = \sum_f \left[ \boldsymbol\alpha \textbf{S}_{f+} \textbf{v}_{f+} \phi_{f+} \right. + \\
& \left.  (1-\boldsymbol\alpha)\textbf{S}_{f-} \textbf{v}_{f-} \phi_{f-}+ \boldsymbol\omega_f(\phi_{f-} + \phi_{f+})  \right]
\end{split}
\label{KNP}
\end{equation}
where $\textbf{S}_{f+}$ is the same as $\textbf{S}_f$ and $\textbf{S}_{f-} = - \textbf{S}_f$. The subscript $f+$ is denoted for the directions coinciding with $\textbf{S}_{f+}$, and $f-$ for the opposite direction. The two terms $ \textbf{S}_{f+} \textbf{v}_{f+} \phi_{f+} $ and $ \textbf{S}_{f-} \textbf{v}_{f-} \phi_{f-} $ in equation \ref{KNP} represent the fluxes evaluated at the $\textbf{S}_{f+}$ and $\textbf{S}_{f-}$ directions respectively. The last part of equation \ref{KNP} represents an additional diffusive term based on the maximum speed of propagation of any discontinuity that may exist at the face. The weighted coefficient $\boldsymbol{\alpha}$ is,
\begin{equation}
\boldsymbol \alpha = \frac{\psi_{f+}}{\psi_{f+} + \psi_{f-}}
\end{equation} 
where $\psi_{f\pm}$ is the local speed of propagation, shown to be:
\begin{align}
\psi_{f+} = max \left( c_{f+} |\textbf{S}_f| + \phi_{f+}, c_{f-} |\textbf{S}_f|+ \phi_{f-}, 0 \right) \\
\psi_{f-} = max \left( c_{f+} |\textbf{S}_f| - \phi_{f+}, c_{f-} |\textbf{S}_f|- \phi_{f-}, 0 \right)
\end{align}
and $c_{f\pm} =\sqrt{\gamma R T_{f\pm}}$ is the local speed of sound at the face. The diffusive volumetric flux $\boldsymbol{\omega}_f$, has the form,
\begin{equation}
\boldsymbol \omega_f = \boldsymbol \alpha (1 - \boldsymbol \alpha) (\psi_{f+} + \psi_{f-}) 
\end{equation}
The scheme implemented to interpolate the values at the center of the face in the directions of  $\textbf{S}_{f+}$ and $\textbf{S}_{f-}$ is based on the limiting standard first and second order upwind \cite{ganapathisubramani2007effects}. The interpolation at $f+$ for example is,
\begin{equation}
\phi_{f+} = (1 - g_{f+}) \phi_O + g_{f+} \phi_N
\end{equation}
where the subscripts $O$ and $N$ represent the nodes at the center of the owner cells and neighbor cells respectively, and the KNP geometric weighting factor $g_{f+}=\beta_f (1- w_f)$ with $\beta_f$ being the van-Leer limiter function. 

All of the gradient terms in the MCT governing equations are computed using the Green-Gauss theorem \cite{patankar1980numerical, moukalled2015finite},
\begin{equation}
\int_{V_c}(\nabla \phi)_c dV=  \sum_f \phi_f \textbf{S}_f 
\end{equation}
where the face value is calculated using the compact stencil method \cite{moukalled2015finite}, which is simply the geometric average of the two cell-centered values of the face,
\begin{equation}
\phi_f = g_c \phi_O + (1- g_c) \phi_N
\end{equation}
where $g_c$ is the geometric weighting factor. The only exception is the pressure gradient, $\hat{\nabla} p$, in the linear momentum equation which was  discretized according to the Kurganov, Noelle and Petrova (KNP) \cite{kurganov2001semidiscrete} flux splitting scheme,
\begin{equation}
\begin{split}
\sum_f \phi_f & \textbf{S}_f  = \sum_f \left[ \boldsymbol\alpha \textbf{S}_{f+} \phi_{f+} +   (1-\boldsymbol\alpha)\textbf{S}_{f-} \phi_{f-} \right]
\end{split}
\end{equation}
where $\alpha$ is the weighted cofficient defined previously.

Finally, the diffusion terms are approximated by,
\begin{equation} 
\int_V \nabla \bigcdot (\Gamma_\phi \nabla \phi) dV = \int_S (\Gamma_\phi \nabla \phi) \bigcdot d \textbf{S} \approx \sum_f (\Gamma_{\phi } \nabla \phi)_f . \textbf{S}_f
 \end{equation}
The  $(\Gamma_\phi \nabla \phi)_f$ term can be obtained as the weighted average of the gradients at the face centroids multiplied by the diffusivity at the centroid,
\begin{equation}
(\Gamma_\phi \nabla \phi)_f = g_c (\Gamma_\phi \nabla \phi)_O + (1-g_c)(\Gamma_\phi \nabla \phi)_N
\end{equation}
In most cases, the diffusivity is interpolated linearly from the cell center values to the faces. The curls of the transport variables are represented by the off diagonal components in the antisymmetric part of the corresponding Green-Gauss gradients. Therefore, the curls of these variables can be computed in a similar fashion to the gradient terms.

Now that the specifics of the finite volume solver have been described, the final step is to give an overview of the algorithm employed. The solver developed is a fully explicit solver: all terms in the MCT governing equations are evaluated at the previous time step. This approach enables fewer computations per time step, but does put a constraint on the size of the time step. The full algorithm of the MCT solver is shown in table \ref{MCT_Alg}. With this algorithm in place, numerical simulation of the compressible flow can be done through the perspective of MCT.

\begin{table}
\caption{Algorithm for solving the MCT governing equations}
 \begin{tabular}{ l}
 \hline
\textbf{while t $<$ End Time:}  \\
\quad Interpolate all the fields from the cell center to face center \\
\quad Calculate the convective, diffusive, and gradient terms\\
\quad Solve the continuity equation for $\rho$ \\
\quad Solve the linear momentum equation for $u_i$ \\
\quad Solve the angular momentum equation for $\omega_i$ \\
\quad Solve the energy equation for $E$ \\
\quad Update the temperature $T$ from $E$\\
\quad Update the pressure using the ideal gas law \\
\quad Update the boundary conditions \\
\quad Update time ($t^{n+1} = t^{n} + \Delta t$)\\
\hline
 \end{tabular}
\label{MCT_Alg}
\end{table}

\section{Verification: \textit{Couette Flow} \label{sec:level4}}

Verification of the compressible MCT solver was done by comparing the numerical results of the compressible isothermal Couette flow with the analytical solution. The assumptions for the Couette flow are that the flow is fully developed, steady state, isothermal, compressible, and two-dimensional \cite{chen2012numerical}, i.e. zero velocity in the y and z direction and zero gyration in the x and y direction. 

\begin{figure}

\centering
\includegraphics [width = 6cm]{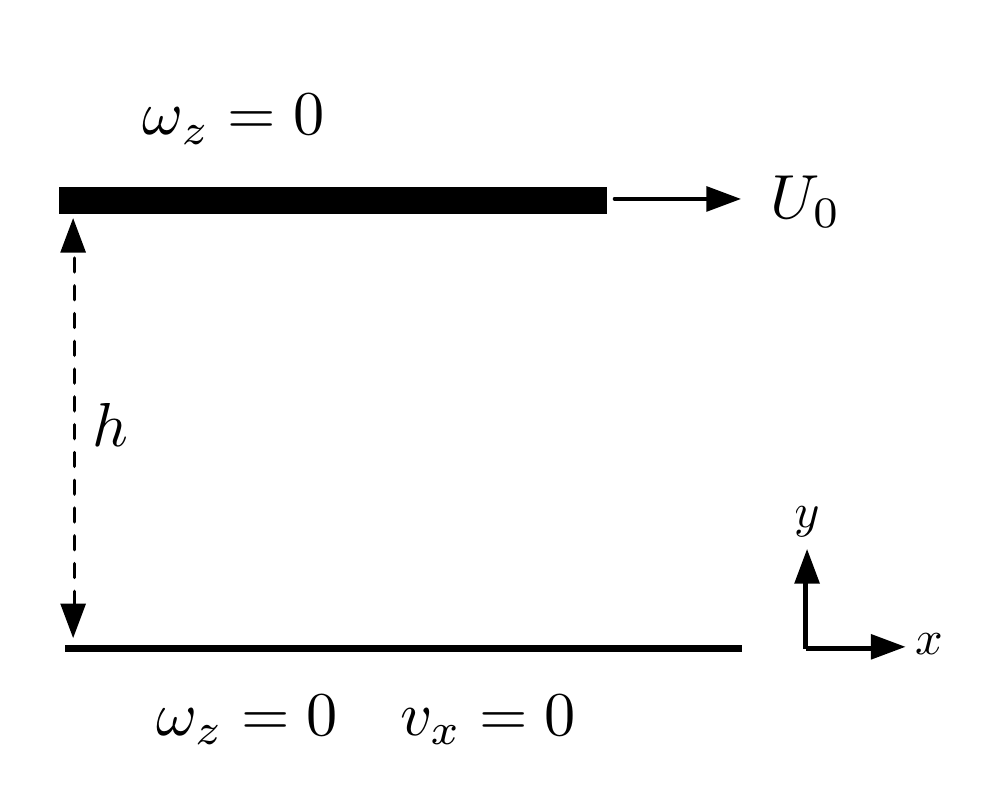}
 \caption{Boundary conditions for a 2D Couette Flow}
 \label{CF}
\end{figure}

Under these assumptions the governing equations for MCT are reduced to:
\begin{align}
(\mu+\kappa) \frac{\partial^2 v_x}{\partial y^2} +\kappa \frac{\partial \omega_z}{\partial y}=0 \\
\gamma \frac{\partial^2 \omega_z}{\partial y^2} - \kappa \frac{\partial v_x}{\partial y} - 2 \kappa \omega_z=0
\end{align}
As for the boundary conditions, the moving plate is placed at a height $h$ above the fixed plate, and moves in the x-direction at the a velocity $U_0$, while the gyration at both plates is fixed at zero due to the no-slip condition. 
Figure \ref{CF} illustrates the boundary conditions of the system. The analytical solutions of gyration and velocity for the Couette flow are,
\begin{align}
\omega_z = C_1 S &\left[ -1+ \left(1-\frac{1}{D}\right)e^{-M y} + \frac{e^{My}}{D} \right] \\
\begin{split}
v_x = C_4 + C_1 &\left[ y + \frac{GS}{M} \left( - \left(1- \frac{1}{D} \right) e^{-My} \right.  \right. \\
 & \left. \left. + \frac{e^{My}}{D} - M y \right) \right]
 \end{split}
\end{align}
where:
\begin{align*}
M &= \sqrt{\frac{\kappa(2\mu+\kappa)}{\gamma(\mu+\kappa)}} ; \quad D = 1 + e^{Mh} ; \quad S = \frac{\kappa + \mu}{\kappa + 2 \mu} \\
G &= - \frac{\kappa}{\kappa + \mu}; \quad C_4 = \frac{ (-2 + D) e^{hM} S G}{F} U_0 ; \\
C_1 &= \frac{D e^{hM} M}{F} U_0; \\
 F &= \left( -1 + e^{hM} \right)^2 GS + D \left[-GS  \right. 
 \\
&\qquad \qquad \qquad \left. +e^{hM}(hM + GS- GhMS) \right]
\end{align*}

\begin{table}
\begin{center}
\caption{\textcolor{black}{Velocity and gyration error analysis}}
 \begin{tabular}{l  c c c c }
 \hline \hline
           & 5x5   & 10x10  & 20x20  &  40x40  \\
 \hline 
 Vel $L_1$ & 0.0265 & 0.0087 & 0.0029 & 0.0012   \\
 Order     & 1.613  & 1.569  & 1.334  &   -       \\ [\smallskipamount]
 Vel $L_2$ & 0.0192 & 0.0060 & 0.0018 & 0.0006   \\
 Order     & 1.681  & 1.717  & 1.632  &   -      \\[\smallskipamount]
 Gyr $L_1$ & 0.0769 & 0.0290 & 0.0091 & 0.0033\\
 Order     & 1.408 & 1.672 & 1.456 & -\\[\smallskipamount]
 Gyr $L_2$ & 0.0265 & 0.0087 & 0.0029 & 0.0012\\
 Order     & 1.579 & 1.769 & 1.693 & - \\[\smallskipamount]
 $\Delta x_{m}$ & 0.2 & 0.1 & 0.05 & 0.025\\
 \hline \hline
 \end{tabular}
\label{table1}
\end{center}
\end{table}

\begin{figure}
\centering
\includegraphics [width = 8cm]{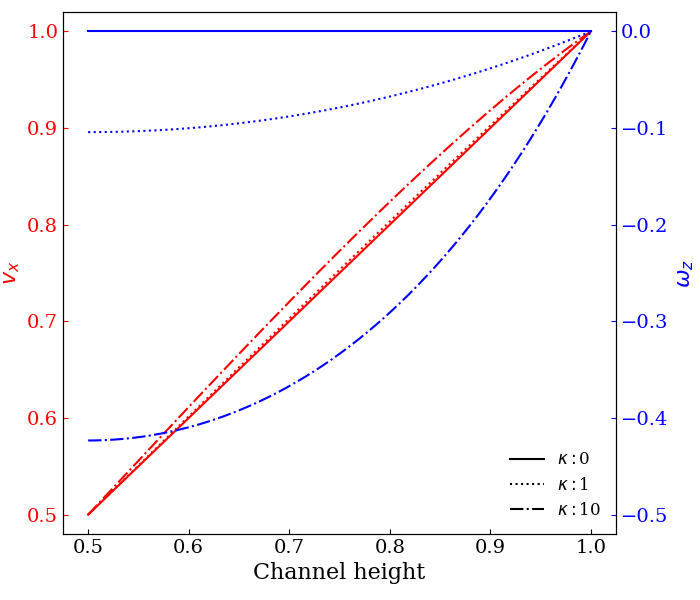}
 \caption{\textcolor{black}{Velocity and gyration profile for the MCT Couette 
Flow}}
 \label{CF2}
\end{figure}

\textcolor{black}{ Figure \ref{CF2} plots the velocity and gyration profiles 
across half of the channel height. Here, the dynamic viscosity, $\mu$, 
and the gyration diffusion coefficient, $\gamma$, were fixed at 1, while the 
value of the rotational viscosity, $\kappa$, varied from 0 to 10. The figure 
shows that, as $\kappa$ increases from 0 to 10, the linearity of the velocity 
profile starts to curve particularly near the boundary. The figure shows that 
the classical rectilinear profile of the Couette flow is a special case of the 
MCT Couette flow at $\kappa=0$. }

The details of the numerical order calculation and verification for the velocity and gyration are shown in Table \ref{table1}. The results clearly indicate that the solver exhibits the desired optimal second order of accuracy.

\section{Validation: \textit{Compression Ramp} \label{sec:level5}}

Finally, the advantage of compressible MCT in capturing the energy cascade at the level of the subscale eddies will be showcased in a shock wave and turbulent boundary layer interaction (STBLI) case, in particular the compression ramp configuration. The compression ramp, has some technical advantages over other STBLI cases, mainly due to the generated shock waves emanating outward through the outflow part of the computational domain, removing the need of imposing a highly accurate far-field boundary condition \cite{adams2000direct}. 

\begin{figure}
  \centering
  \includegraphics[width = 7cm]{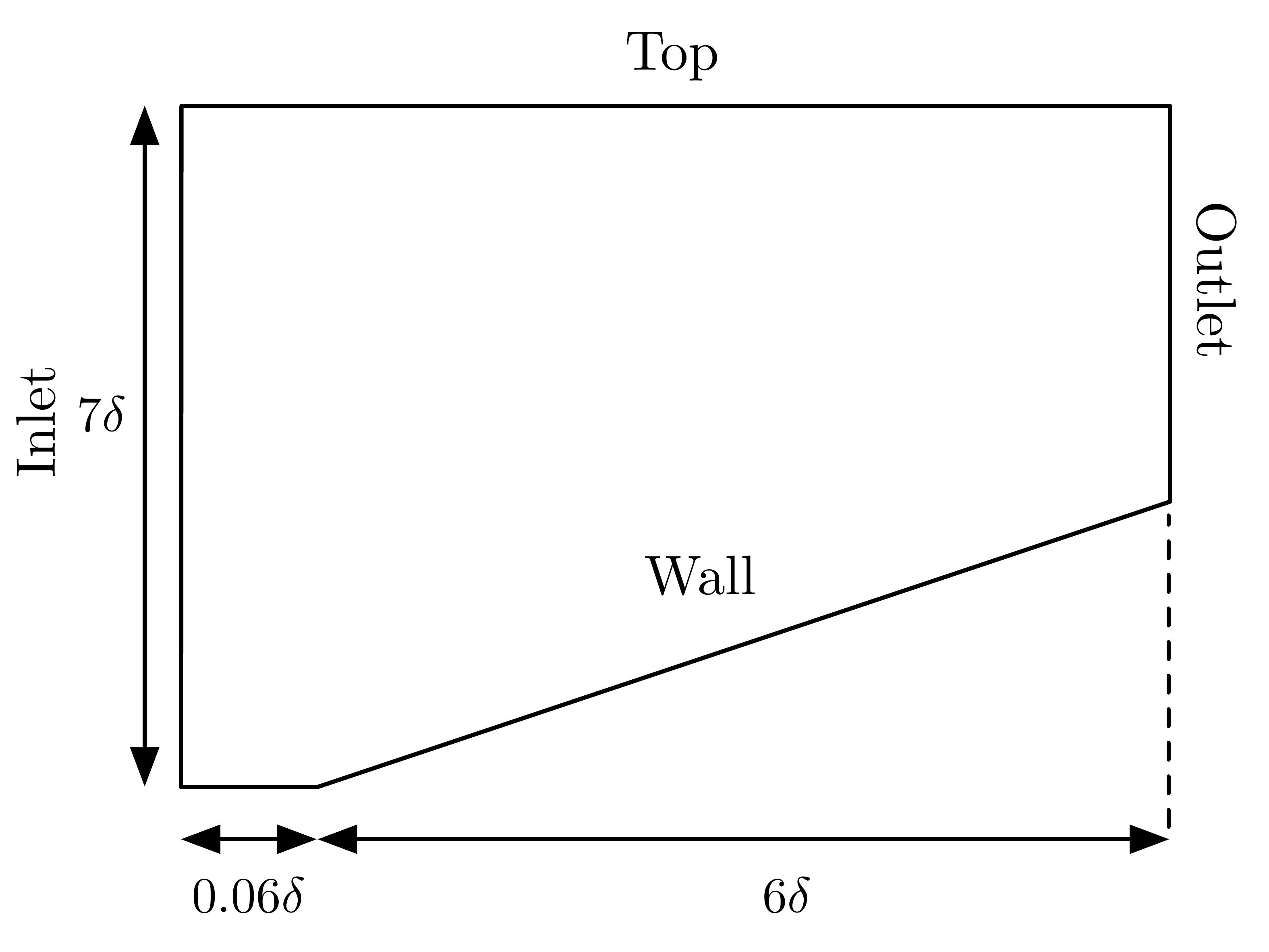}
  \caption{Size of the \textcolor{black}{compression ramp} computational domain}
  \label{CR}
\end{figure}

For our particular case, Kuntz et. al.'s experiment \cite{kuntz1985experimental, kuntz1987turbulent} of a supersonic flow over an 
8$^{\circ}$ compression ramp is replicated. In his paper Kuntz et. al. 
considered a series of five compression ramps ranging from 8$^{\circ}$ to 
24$^{\circ}$. Using this set of ramp angles Kuntz was able to capture a full 
range of possible flow fields, including flow with no separation, flow with 
incipient separation, and flow with a significant amount of separation. 
\textcolor{black}{Kuntz et. al.'s experimental data has been referenced to 
derive shock-wave/boundary-layer interaction (SWBLI) models based on mass 
conservation \cite{souverein2013scaling}. In addition, this data was used to 
validate the accuracy of different RANS models 
\cite{oliver2007assessment,asmelash2013numerical}, to analyze the 
significance of the spanwise geometry variation and to relate it to a canonical 
compression flow for a three-dimensional bump flowfield 
\cite{tillotson2009experimental}.} For the 8$^{\circ}$ compression ramp, Kuntz's 
experimental results showed no separation of the flow near the corner ramp, 
making it an ideal simple case to demonstrate the capabilities of MCT.  Another 
reason why the 8$^{\circ}$  compression ramp is chosen is the two-dimensional 
behavior of the shock near the ramp corner, giving credence to the assumption of 
a two-dimensional flow, as well as the adiabatic condition at the wall, 
resulting in no heat dissipation \cite{bernardini2016heat}. Figure \ref{CR} 
shows a schematic for the present ramp configuration. 

\subsection{Material Parameters}
The working fluid is assumed to be an ideal gas, where the equation of state is $p = \rho RT$. The gas constant is taken as $R = 287.06$ $m^2s^{-2}K^{-1}$, the specific heat coefficient for constant pressure is $c_p = 1004.06$ $J/(kgK)$ and the Prandtl number is $Pr = 0.7$. The summation of all the viscous coefficients were computed by Sutherland's law:
\begin{align}
\kappa + \mu = \frac{(1.458 \times 10^{-6}) T^{3/2}}{T +110.4}
\end{align}
The incoming freestream conditions are listed in Table \ref{table2} as reported in the experiment of Kuntz et. al. \cite{kuntz1987turbulent}.
\begin{table}
\begin{center}
\caption{Freestream flow conditions taken form the experiment setup of Kuntz et. al. \cite{kuntz1985experimental}}
 \begin{tabular}{r r r r}
$p_\infty$ [Pa] & $\rho_\infty$ [kg/m$^{\text{3}}$]& $U_\infty$ [m/s] & $T_\infty$ [K] \\
\hline
14319 & 0.465 & 612 & 107.79\\
 \end{tabular}
\label{table2}
\end{center}
\end{table}
The temperature at the wall was set to adiabatic conditions, in reference to the experiments by Kuntz et. al. \cite{kuntz1985experimental}. The boundary layer thickness, $\delta$, and the momentum layer thickness, $\theta$, for the incoming flow, reported by Kuntz et. al. \cite{kuntz1985experimental}, at the location of the ramp edge were measured to be 8.27 mm and 0.57 mm. As for the MCT variables,  Wonnell and Chen \cite{wonnell2017morphing} showed that the viscous forces arising from the gyration should be around 99 times  the dynamic viscosity (i.e. $\kappa = 99 \mu$) to obtain a turbulent incompressible flow. This study follows the work of Wonnell and Chen  by making $\kappa$ equal to $99 \mu$ \cite{wonnell2017morphing}. The two other dimensionless parameters ($C_\alpha$ and $C_\beta$) are set to zero, since currently there is no physical meaning to them. Table \ref{table3} shows the MCT dimensionless parameters introduced in section \ref{sec:level2} computed from the freestrean conditions, and the length scale parameter $L=\sqrt{j}$.

\begin{table}
\begin{center}
\caption{Dimensionless numbers based of the freestream conditions of the experimental setup}
 \begin{tabular}{l c c c c c c c c c c c c c c}
  $j$ [m$^{\text{2}}$] & $Ma$ & $Re$ & $Er$ & $C_\gamma$ & $C_\alpha$ & $C_\beta$   \\
\hline
 10$^{\text{-6}}$ &2.94 & 38000 & 38400 & 1.5809 $\times$ 10$^{\text{5}}$  & 0 & 0
 \end{tabular}
\label{table3}
\end{center}
\end{table}

\subsection{Boundary Conditions and Meshes}

The subject of spatially evolving turbulent flows poses a particular challenge for numerical simulation, due to the need for time-dependent inlet conditions at the upstream boundary. In many cases, the downstream flow is highly dependent on the conditions of the inlet. Therefore it is necessary to specify a realistic time series of turbulent fluctuations that are in equilibrium with the mean flow, while still satisfying the governing equations. For this reason, creating accurate inflow turbulent conditions may require costly independent simulations \cite{martin2007direct}, forced transition \cite{rist1995direct}, a long leading edge  \cite{oliver2007assessment}, or cost-saving but crude inflow generation methods \cite{xu2004assessment}. 

Oliver tested turbulent RANS models for a flow past an $8^{\circ}$ compression ramp \cite{oliver2007assessment}. In this study, the length of the flat plate upstream of the ramp corner exceeded $60\delta$. The reason for this addition was to allow the inflow to develop from a uniform to a turbulent flow, with a boundary layer that matched the experimental boundary layer thickness. 

Here, MCT has the ability to control the eddy structure of the flow by the gyration term, enabling it to model turbulence without the need for complex boundary conditions. Wonnell and Chen \cite{wonnell2017morphing} showed through utilizing the subscale eddies near the wall that MCT can control the regime of the flow and change it from laminar to transitional or turbulent. They  later showed that in addition to controlling the eddies near the wall, one can control the eddies' rotational speed at the inlet, and thus control the incoming turbulent kinetic energy $1/2 \rho j \omega_k \omega_k$ \cite{wonnell2016morphing2}. 

\begin{figure}
\centering
\includegraphics[width=8cm]{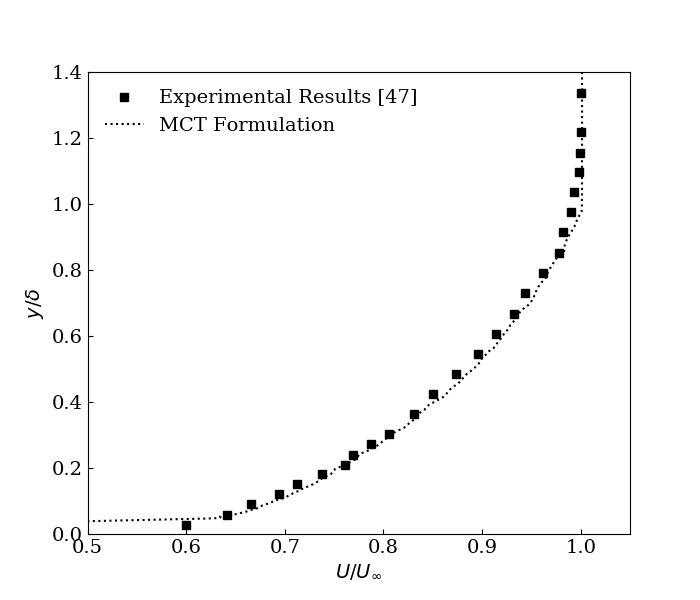}   
\caption{Velocity profile of the incoming flow implemented in both MCT cases versus the experimental data \cite{kuntz1985experimental}}
\label{Vel_In}
\end{figure}

 The inflow varaibles implemented in the current case to achieve a turbulent flow are decomposed into two parts, the mean and fluctuating components. For the mean flow, a prescribed turbulent mean velocity profile was defined at the inlet, through the implementation of Martin's procedure \cite{martin2007direct}. Figure \ref{Vel_In} plots the inlet velocity profile from the MCT simulation with the experimental incoming velocity profile \cite{kuntz1987turbulent} located $0.06\delta$ upstream of the compression ramp.  

\begin{figure}
  \centering
  \includegraphics[width=8cm]{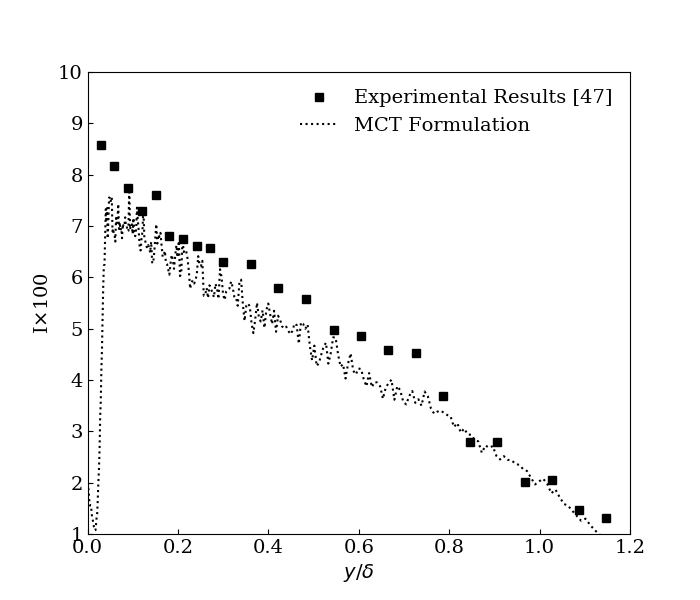}
  \caption{Turbulent Intensity at the inlet for  MCT and experimental results}
  \label{TI}
\end{figure}

The fluctuations are generated by controlling the rotational speed of the upstream eddies. This happens because the instantaneous inlet gyration $\omega_k$ is decomposed into mean and fluctuating parts:
\begin{equation}
\omega_k(t,y) = \langle \omega_k (y)\rangle + \omega_k'(t,y) 
\end{equation}
where  $\langle \omega_k \rangle$ is the mean value of the gyration, and $ \omega_k'(t)$ is the fluctuating rotation speed of the eddy. The perturbations are produced through a random number generator with the range of values constrained by the root-mean squared (rms) gyration, and turbulent intensity from the experiments at the specified point. The rms value of the perturbed gyration becomes
\begin{equation}
\omega_{rms} = \sqrt{\frac{1}{N} \sum_{k=1}^N \omega_i' \omega_i'}
\end{equation}
and  the turbulent intensity of the MCT flow becomes
\begin{equation}
I = \frac{\omega_{rms}d/2}{U_\infty}
\end{equation}
It can be seen that the larger the range of the perturbation in the gyration field the larger the rms value and thus the larger the turbulent intensity. In order to focus on the effects of the fluctuations, the mean gyration was set to zero, while the amplitude of the perturbed gyration was defined so that the turbulent intensity of the incoming flow matches the experimental turbulent intensity results of Kuntz  et. al.  \cite{kuntz1985experimental} as shown in Figure \ref{TI}.

The remaing boundary conditions at the inlet are the pressure and temperature, which are set to the freestream conditions in Table \ref{table2}. At the outlet and top boundaries, supersonic outflow boundary conditions are implemented, and for the ramp wall the no-slip and adiabatic boundary conditions are implemented.

\begin{figure}
\includegraphics[width=8cm]{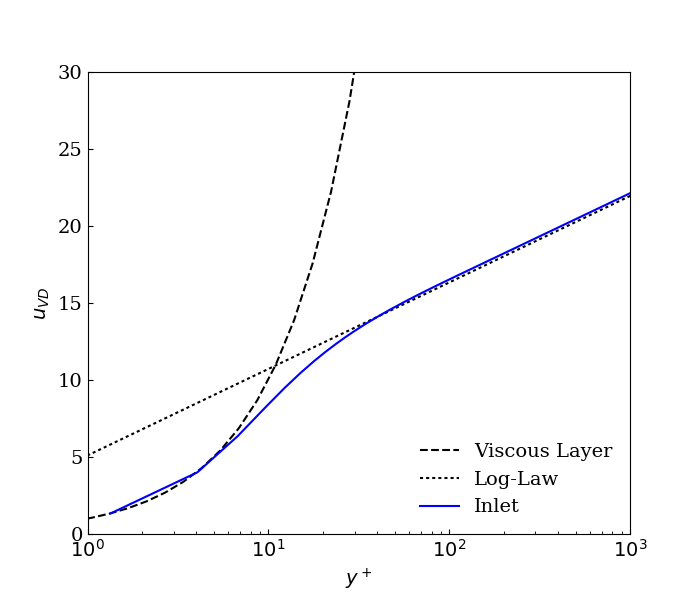}  
\caption{ Van Driest transformed streamwise velocity profile at the inlet}
\label{Van}
\end{figure}

A structured grid is generated, with the distance between the corner and the outlet equal to $6\delta$, and the length upstream of the corner equal to $0.06\delta$. The number of cells used in the current simulation is $505$  in the streamwise and $1000$ in the wall-normal directions. In the wall-normal direction, the grid spacing near the wall is $\Delta y^{+} = 1.34$ with $10$ grid points within $y^+ <30$. Figure \ref{Van} plots the Van Driest transformed velocity at the inlet. It is evident from the figure that the 
cell resolution in the y-direction is sufficient to capture the viscous sublayer and the logarithmic region of the velocity profile.

\subsection{Comparison between the Simulation and Experiments}
\begin{figure}
  \centering
  \includegraphics[width=8cm]{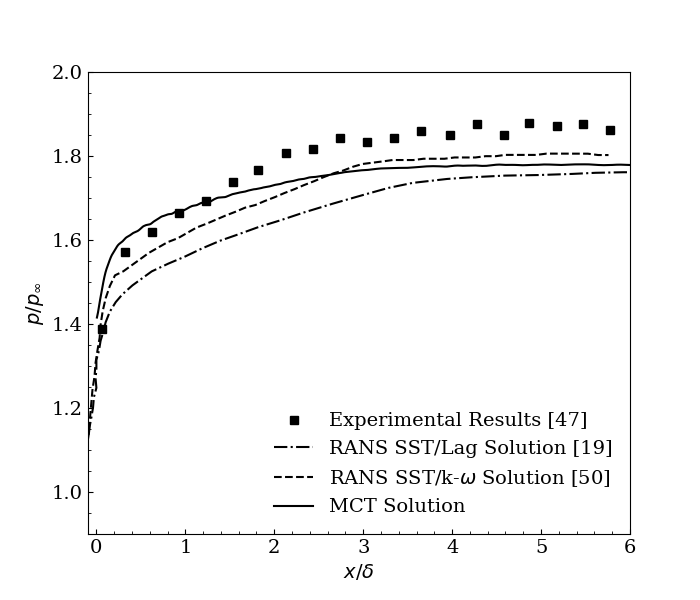}
  \caption{Mean wall pressure distribution from MCT and experimental results}
  \label{WP}
\end{figure}

Validation of the proposed MCT scheme was conducted through comparing the pressure at the wall as well as the velocity profile between the experiments and the simulation. Figure \ref{WP} plots the normalized wall pressure of the experimental results versus the RANS results of Oliver \cite{oliver2007assessment} \textcolor{black}{and Asmelash \cite{asmelash2013numerical}}, and the proposed MCT numerical solver results. 

The figure shows that the MCT solution comes closer to predicting the 
experimental wall pressure than the turbulent RANS models, especially near the 
ramp edge where MCT captured the first four points of the experimental data 
while RANS only captured the first point. The difference between the RANS and 
MCT wall pressure results can be attributed to the convective scheme implemented 
in each case. In the RANS simulations of the compression ramp, Oliver 
\cite{oliver2007assessment} implemented a first order upwind scheme, 
\textcolor{black}{and Asmelash \cite{asmelash2013numerical} implemented a a 
second order upwind scheme. Here, the MCT scheme is a second-order 
generalization of the Lax-Friedrichs scheme.} It is also worthwhile to mention 
that the mesh requirement for the MCT case is less demanding compared with a 
similar DNS study for a compression ramp \cite{wu2007direct}. 
\begin{figure}
\begin{subfigmatrix}{2} 
\centering
 \subfigure[Velocity Profile at $3\delta$] { \includegraphics[width=8cm]{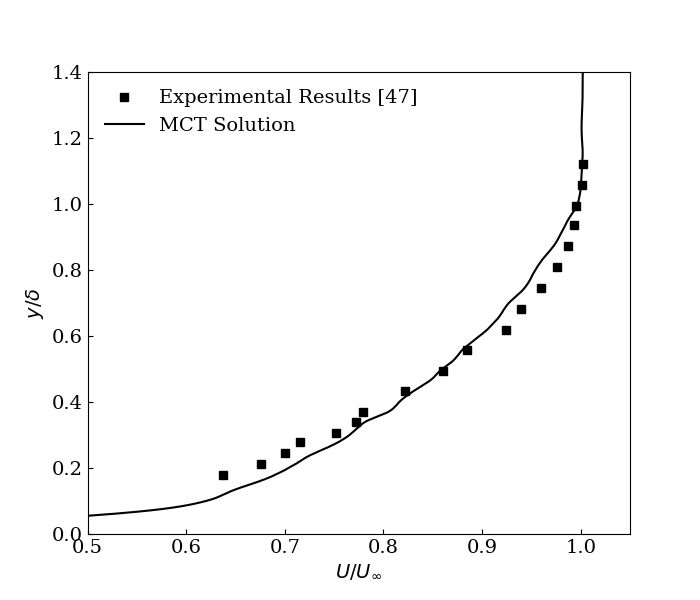}}
 \subfigure[Velocity Profile at $4.2\delta$]  { \includegraphics[width=8cm]{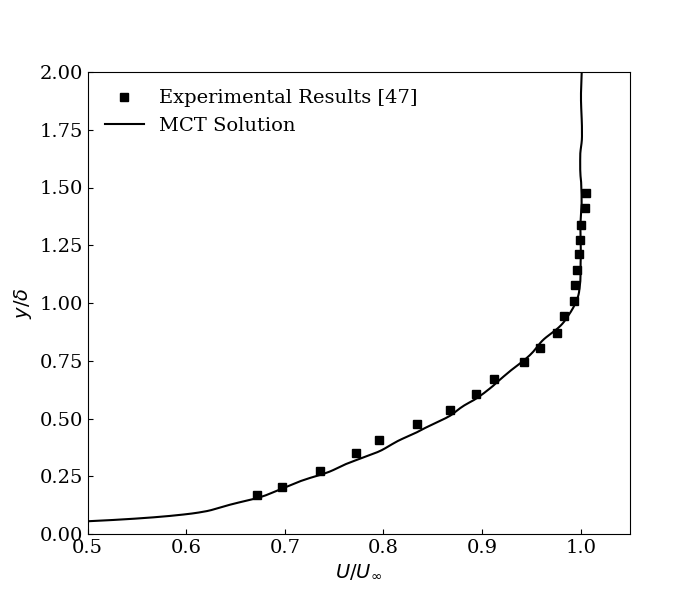}}
  \end{subfigmatrix}
  \begin{center}
   \subfigure[Velocity Profile at $5.4\delta$]  { \includegraphics[width=8cm]{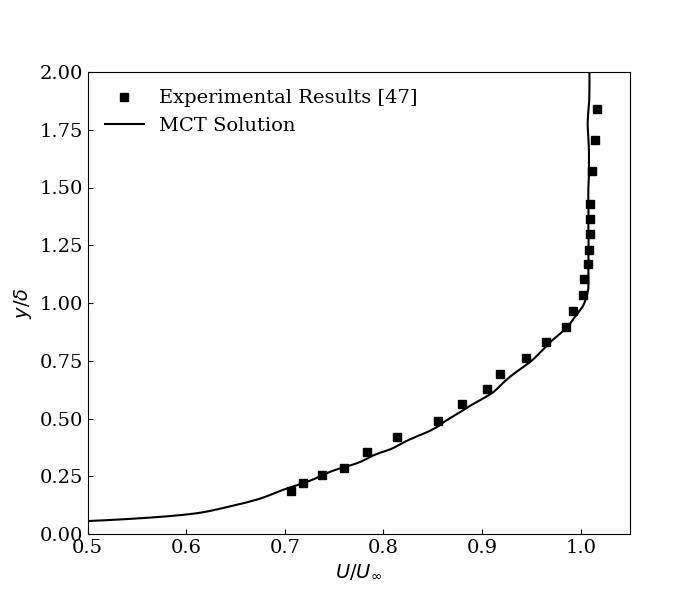}}
  \end{center}
\caption{Velocity profiles at (a) $3\delta$, (b) $4.2\delta$, and (c) $5.4\delta$ downstream of the ramp edge from the MCT solution and the experiements}
\label{Vel_dw}
\end{figure}
The grid spacing near the wall for the MCT case is $\Delta y^{+} = 1.34$ with 
$10$ grid points for $y^+ <30$, while for a similar DNS simulation 
\cite{wu2007direct}, the required spacing normal to the wall is $\Delta y^{+} = 
0.2$ with more than 20 grid points in  $y^+ <20$. Unlike the classical DNS 
relying on fine meshes to resolve subscale motions, MCT formulates subscale 
motions into the governing equations. Therefore, the mesh requirements for MCT 
are less restrictive than DNS, resulting MCT as a more computation-friendly 
theory for turbulent flows. Figure \ref{Vel_dw} shows the normalized flow 
velocities at three locations $3\delta$, $4.2\delta$, and $5.4\delta$ downstream 
from the ramp corner, and the MCT numerical solver results. The figure shows 
that MCT is capable of capturing the boundary layer profile inside the shock.
\subsection{Subscale Kinetic Energy }
As stated previously, the aim of this paper is to investigate the energy transfer between the subscale eddies and the bulk flow inside the shock. Chen \citep{chen2012numerical} stated that the total energy density of each subscale eddy can be expressed as,
\begin{equation}
E = \frac{1}{2}\left(u_iu_i + j \omega_i\omega_i \right)+ e
\label{E_TOT}
\end{equation}
where $\frac{1}{2}u_iu_i$ contributes to the translational kinetic energy, $\frac{1}{2}j \omega_i\omega_i$ contributes to the rotational kinetic, and $e = c_v T$ represents the internal energy density of the flow. Analysis of the energy cascade is acheived by the use of the conventional Reynolds averaging (also known as time averaging) method and the mass-weighted averaging method or better known as Favre averaging. The main advantage of these methods is in their ability to resolve the relevant physical processes at different scales \cite{guarini_moser_shariff_wray_2000}. The following notations are used for the mean values: $\langle$ $\rangle$ for the Reynolds average and $\{$ $\}$ for the Favre average, which is defined as  
\begin{align*}
\{ \phi \} = \frac{\langle \rho \phi  \rangle}{ \langle \rho \rangle}
\end{align*}
where $\phi$ represent any time-dependent variable. Here, the single prime represents the Reynolds fluctuation, and double prime represents the Favre fluctuations. 

The scale decomposition employed in the total energy density equation \ref{E_TOT} is carried out using Favre filtering in order to account for the density fluctuations of the flow. The Favre decomposition of the total energy density is,
\begin{equation}
\begin{split}
E =\frac{1}{2}&\{u_i\}\{u_i\} + \{u_i\}u_i''  + \frac{1}{2}u_i''u_i'' + \frac{j}{2}  \{\omega_i\}\{\omega_i\} 
\\ &+ j\{\omega_i \}\omega_i''  + \frac{j}{2}\omega_i''\omega_i''+ \{e\} + e''
\end{split}
\end{equation}
The first term on the right hand side $\frac{1}{2}\{u_i\}\{u_i\}$ represents the Favre-averaged mean flow translational kinetic energy, and represents the mean translational speed of the flow. The second term satisfies the relation $\langle \rho \{u_i\}u_i'' \rangle = 0$ and may be called the Favre-fluctuating mean flow translational kinetic energy. Huang \cite{huang1995compressible} gives a physical interpretation to the second term by examining the turbulent diffusion in the total energy equation. The final term corresponding to the translational motion is $\frac{1}{2}u_i''u_i''$, and refers to the Favre-fluctuating translational kinetic energy. Similarly, one may define the rotational components of the kinetic energy, the Favre-averaged mean flow rotational kinetic energy as $\frac{j}{2}  \{\omega_i\} \{\omega_i\}$, the Favre-fluctuating mean flow rotational kinetic energy as $j\{\omega_i\}\omega_i''$, and the Favre-fluctuating rotational kinetic energy as $\frac{j}{2}\omega_i''\omega_i''$. Finally, $\{e\}$ is the Favre-averaged internal energy, and $e''$ is the Favre-fluctuating internal energy. 

\begin{figure}
  \centering
  \includegraphics[width=8cm]{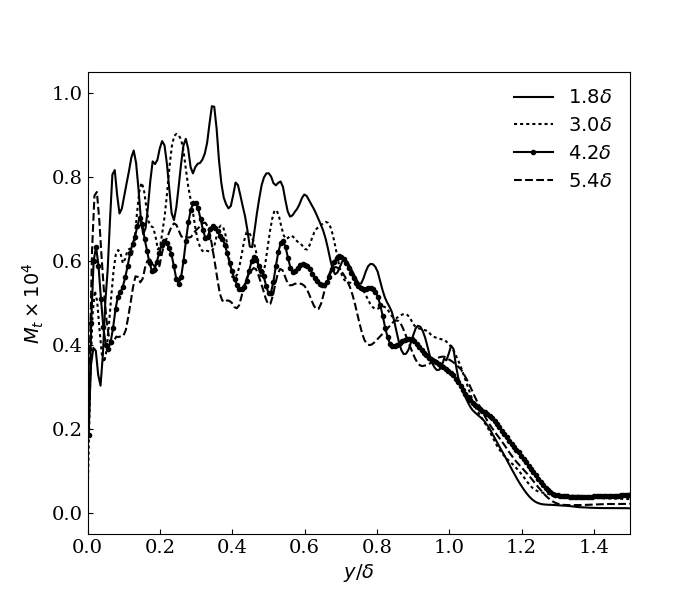}
  \caption{Turbulent Mach number at different location at (a) $1.8\delta$, (b) $3\delta$, (c) $4.2\delta$, and (d) $5.4\delta$ along the ramp}
  \label{Mt}
\end{figure}

Applying Reynolds averaging over the Favre-decomposed total energy density yields the mean component of the total energy density,
\begin{align}
\begin{split}
\langle E \rangle &= \{u_i\} \left( \langle u_i\rangle -\frac{\{u_i\}}{2} \right) + j\{\omega_i\} \left( \langle \omega_i \rangle - \frac{\{\omega_i\} }{2}  \right) 
\\ & + \frac{1}{2}\langle u_i''u_i'' \rangle  + \frac{j}{2} \langle \omega_i''\omega_i'' \rangle+ \{e\} + \langle e'' \rangle
\end{split}
\label{K_Rey}
\end{align}
The first two terms on the right hand side represent the contribution of the mean translational and mean rotational kinetic energies to the mean total energy density. The next two terms represent the contribution of the averaged Favre-fluctuations to the mean total  energy. The $\frac{1}{2}\langle u_i''u_i'' \rangle$ term is found in most classical papers discussing turbulence, and is used in the computation of the turbulent Mach number. The other term $\frac{j}{2} \langle \omega_i''\omega_i'' \rangle$ is strictly unique to an MCT flow, and represents the fluctuations in the subscale eddies' rotational speed. Therefore, an MCT flow adds to the classical turbulent Mach number a component from the eddies' rotation,
\begin{align}
M_t = \frac{\sqrt{\frac{1}{2}\langle u_i'' u_i'' \rangle+\frac{j}{2} \langle \omega_i''\omega_i'' \rangle}}{\langle c \rangle}
\end{align}
where $\langle c \rangle$ represents the Reynolds average speed of sound. Figure \ref{Mt} plots the turbulent Mach number for the 8 degree compression ramp at different locations along the streamwise direction. For locations near the ramp edge the turbulent Mach number is higher than it is further downstream. The explanation for the decay in the fluctuations will be given in the following part of the discussion.
The last two terms in Eq. \ref{K_Rey} represent the contribution of the mean Favre internal energy, and the average Favre-fluctuating internal energy to the mean total energy density. Note that $\{e\} + \langle e'' \rangle = \langle e \rangle$. The reason the mean Reynolds internal energy is not represented is to see the contribution of the Favre fluctuations to the flow.
\begin{figure*}
  \centering
  \includegraphics[width=\textwidth,height=9cm]{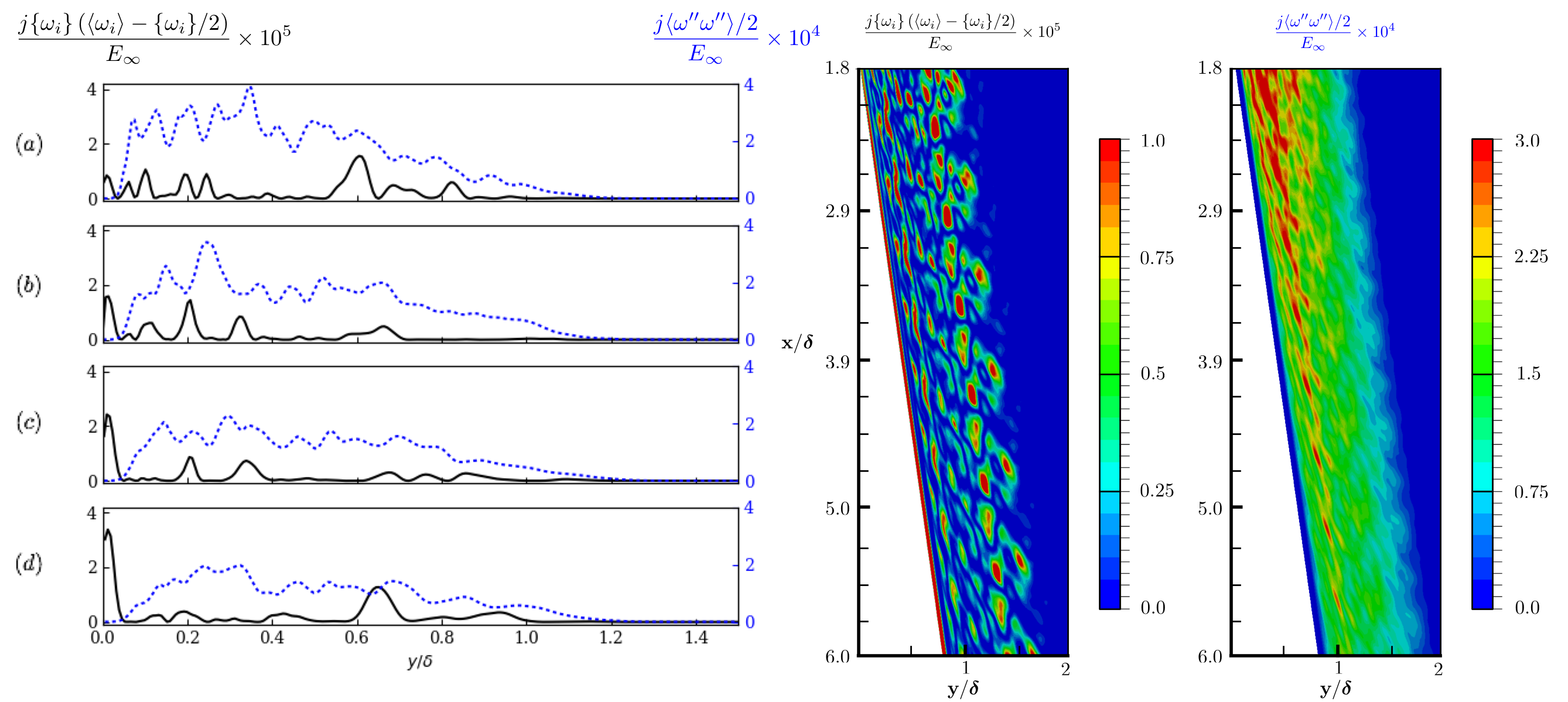}
  \caption{Rotational kinetic energy component of the mean total energy density, mean (\sampleline{}) and fluctations (\sampleline{dashed}), at (a) $1.8\delta$, (b) $3\delta$, (c) $4.2\delta$, and (d) $5.4\delta$ along the ramp}
  \label{Mean_KE_Rot}
\end{figure*} 
\begin{figure*}
  \centering
  \includegraphics[width=\textwidth,height=9cm]{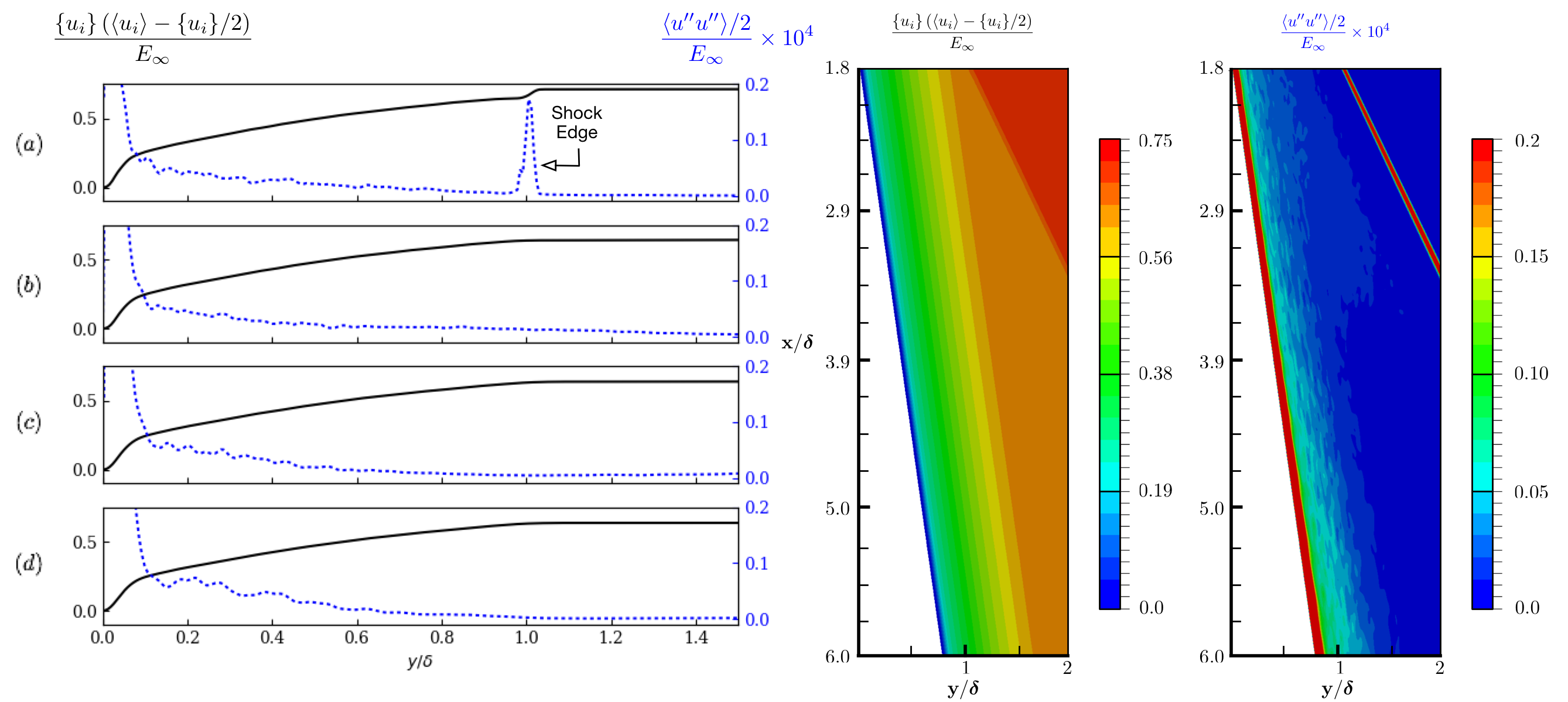}
\caption{Translation kinetic energy component of the mean total energy density,  mean (\sampleline{}) and fluctations (\sampleline{dashed}), at (a) $1.8\delta$, (b) $3\delta$, (c) $4.2\delta$, and (d) $5.4\delta$ along the ramp}
  \label{Mean_KE_Trans}
\end{figure*}

\begin{figure*}
  \centering
  \includegraphics[width=\textwidth,height=9cm]{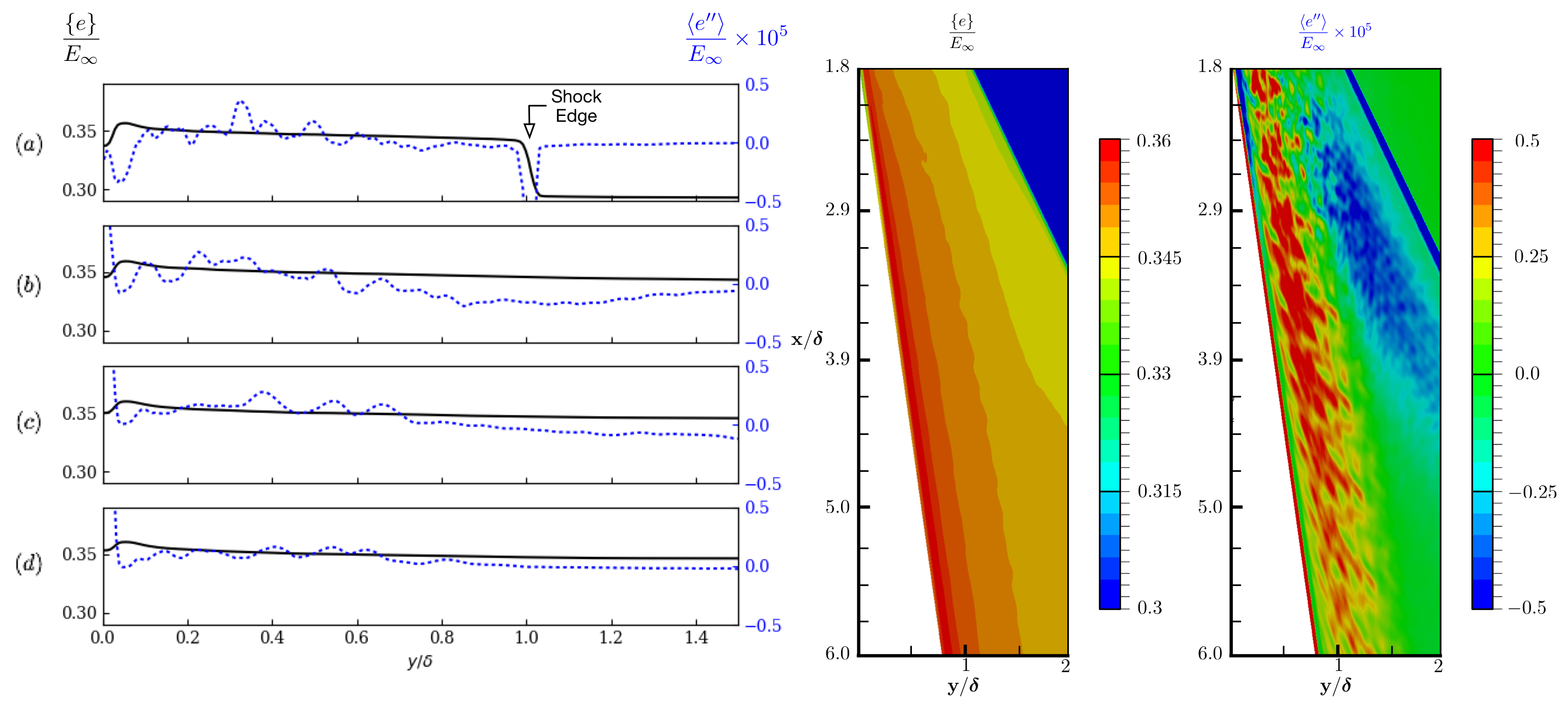}
  \caption{Internal energy component of the mean total energy, mean (\sampleline{}) and fluctations (\sampleline{dashed}), at (a) $1.8\delta$, (b) $3\delta$, (c) $4.2\delta$, and (d) $5.4\delta$ along the ramp}
  \label{Mean_KE_Int}
\end{figure*} 

In order to understand the energy cascade at the level of the subscale eddies, the rotational component of the mean total energy density is investigated. Figure \ref{Mean_KE_Rot} compares the mean rotational component $ j \{\omega_i\} \left( \langle \omega_i \rangle - \frac{\{\omega_i\} }{2}  \right) $ with the averaged Favre-fluctuating rotational component $ \frac{j}{2} \langle \omega_i''\omega_i''\rangle$ at different locations along the ramp. The variables were normalized with respect to the the freestream total energy density $E_\infty = \frac{1}{2} U_\infty^2 + c_v T_\infty$. The figure clearly shows that the averaged component of the rotational kinetic energy density is zero outside the boundary layer indicating an irrotational bulk flow, as was specified at the inlet boundary ($\langle \omega_i \rangle_{inlet} = 0$). Near the wall ($y/\delta < 0.1$), an increase in the magnitude of the averaged component of the rotational kinetic energy density is clearly observed, which can be attributed to the shear forces arising from the wall as well as the diffusion of the near-wall eddies as is clear in the contour plot of Figure \ref{Mean_KE_Rot}. Inside the boundary layer but away from the wall ($0.1<y/\delta < 1$), the figure shows areas with large values of mean rotational kinetic energy, indicating the presence of eddies. It can be seen from the figure that the eddies near the boundary layer are more are tightly packed then the eddies near the walls which are more stretched and elongated. The fluctuating component of the rotational kinetic energy starts out with a large magnitude and decays as it moves along the ramp to less than a half. The reason for having large values of the fluctuation near the ramp edge is due to their proximity to the inlet, which has a boundary condition to generate turbulence by adding fluctuations to the rotational speed of the flow ($\langle \omega'_i\omega'_i\rangle_{inlet}   = \omega_{rms}^2$). Moreover, the profile of the fluctuations at $x/\delta = 1.8$ is consistent with the inlet condition, since the turbulent rotational speed is defined inside the boundary layer and diminishes at the edge of the boundary layer. Finally, when comparing the fluctuations along the ramp, the plot shows a large number of local minima and maxima near the ramp edge, with rapid variation between each extremum. This trend implies that there are a lot of small subscale eddies, each separate from the other, as is clear in the contour plot of $\frac{j}{2} \langle \omega''_i\omega''_i \rangle$. Further along the ramp, the plot for $ \frac{j}{2} \langle \omega''_i \omega''_i\rangle$ shows fewer local minima and maxima with a slower rate of change for each extremum. The results imply that a lot of the previous small subscale eddies merge together or diffuse into the mean flow. This behavior is evident from the contour of $\frac{j}{2} \langle \omega''_i\omega''_i \rangle$. The impact of rotational kinetic energy on the translational kinetic energy and internal energy will be shown in the following discussions.

 Starting with the translational kinetic energy, figure \ref{Mean_KE_Trans} plots the normalized mean components of the translational kinetic energy $\{u_i\} \left( \langle u_i\rangle -\frac{\{u_i\}}{2} \right)$ as well as the normalized averaged Favre-fluctuating components $\frac{1}{2}\langle u_i''u_i'' \rangle$ at different locations along the ramp. It can be seen from the figure that the biggest contributor to the total energy is the mean translational kinetic energy component of the flow, with the averaged Favre-fluctuations component being smaller than the freestream total energy by four orders of magnitude. The behavior of the averaged Favre-fluctuations translational kinetic energy is decomposed into the near-wall section ($y/\delta < 0.1$), and the boundary layer section ($0.1 < y/\delta < 1$). For the near-wall part, an increase in the magnitude is observed  along the streamwise direction. This increase is highly associated with the shear forces arising from the wall, as well as the increase in the rotational speed of the subscale eddies near the wall. The boundary layer section shows an increase in the average Favre-fluctuating translational kinetic energy along the ramp, coinciding with the decrease of the average Favre-fluctuating rotational component of the flow. In summary, the eddies' rotational energy is dissipated into the translational fluctuating energy.

The other aspect of this energy transfer involves the transmission of rotational kinetic energy to internal energy. Figure \ref{Mean_KE_Int} compares the Favre-averaged internal energy $\{ e \}$ with the averaged Favre-fluctuating internal energy $\langle e'' \rangle$ at different locations along the ramp. From the figure, it is evident that the mean component of the internal energy is constant except near the wall where it is increasing in magnitude along the streamwise direction. The averaged Favre-fluctuating internal energy, away from the wall starts with a maximum value of 0.4 and decreases along the streamwise direction. The large value near the ramp edge, and the large oscillations in the averaged Favre-fluctuating internal energy is directly related to the rotational speed of the subscale eddies, and in particular the averged Favre-fluctuating rotational component of the total energy density. When the averged Favre-fluctuating rotational kinetic energy component of the total energy is high, this increase in turn leads to high fluctutations in the averaged Favre-fluctuating internal energy $\langle e'' \rangle$, as the averged Favre-fluctuating rotational kinetic energy decays along the ramp so does the averaged Favre-fluctuating internal energy. One can conclude that the fluctuations in the internal energy are created from the fluctuations in rotational kinetic energy. Still as the eddies move along the streamwise direction, they diffuse and merge with the mean component of the energy, resulting in a decay in the average fluctuating component of the internal energy. Figure \ref{Mean_KE_Int} clearly confirms that along the streamwise direction a decay in the fluctuating component of the internal energy occurs.

\begin{figure}
  \centering
  \includegraphics[width=\textwidth /2]{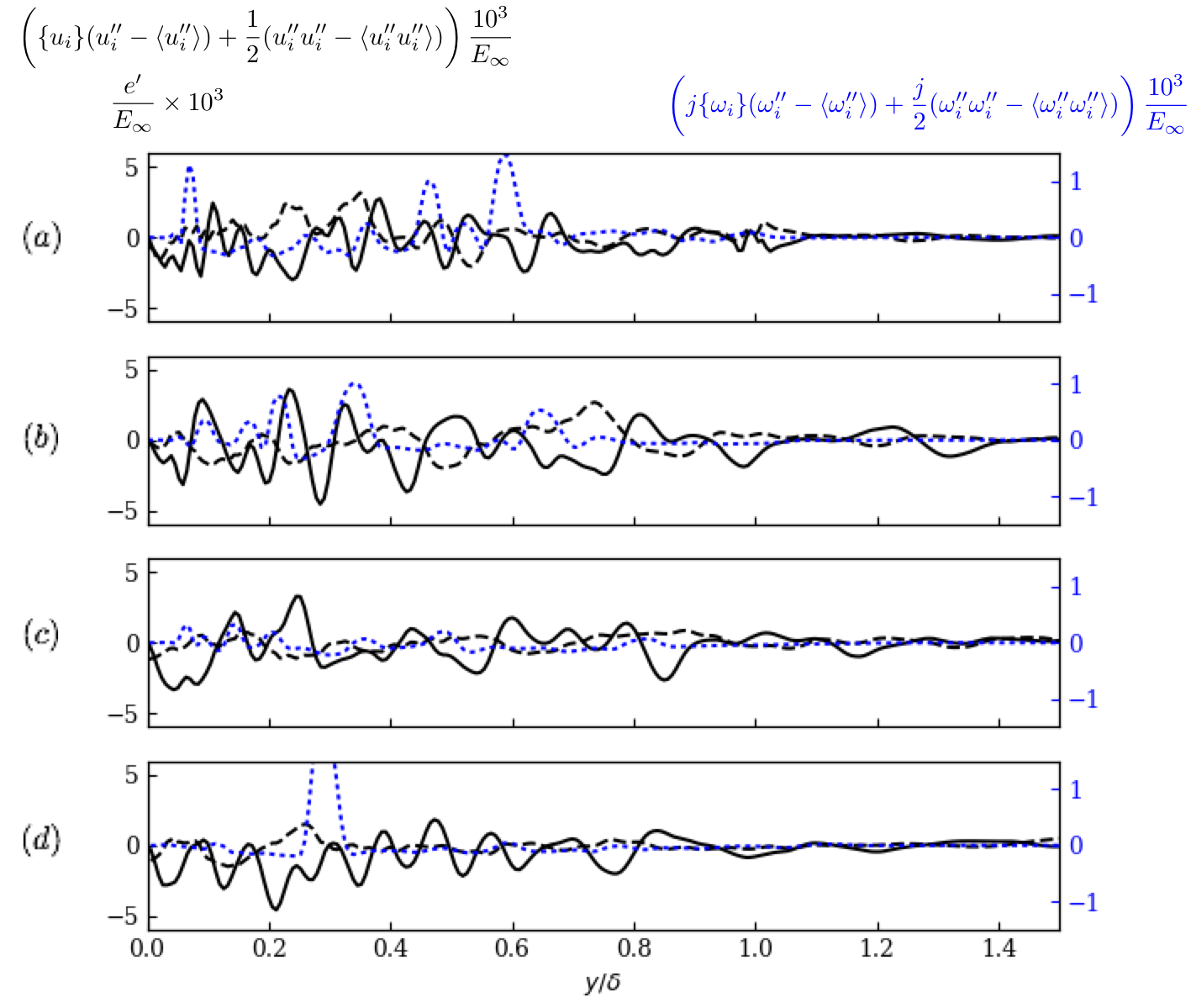}
  \caption{Instantaneous total energy density profiles, translational component (\sampleline{}), internal component (\sampleline{dashed}) and rotational (\textcolor{blue}{\sampleline{dotted}}), at (a) $1.8\delta$, (b) $3\delta$, (c) $4.2\delta$, and (d) $5.4\delta$ along the ramp}
  \label{Fluct_KE}
\end{figure}

The final component of the total energy density is the instantaneous fluctuating part, 
\begin{align}
\begin{split}
E' & = \{u_i\}\left( u_i'' -\langle u_i '' \rangle \right) + \frac{1}{2}(u_i''u_i'' - \langle u_i''u_i'' \rangle) 
\\ & + j\{\omega_i\} \left( \omega_i'' - \langle \omega_i \rangle \right) + \frac{j}{2}(\omega_i''\omega_i'' - \langle \omega_i''\omega_i'' \rangle) 
+ e'
\end{split}
\label{k_p}
\end{align}
Figure \ref{Fluct_KE} plots the translational kinetic energy component of 
equation \ref{k_p}, as well as the internal energy and rotational kinetic energy 
components at different locations along the ramp at a time step t = 0.005 
seconds. The figure clearly shows that the fluctuations in the eddies' 
rotational kinetic energy at the inlet had an effect on the instantanous 
fluctutations in the translational kinetic energy as well as the internal energy 
of the eddies.\\

\section{Conclusion \label{sec:level6}} 

A shock-preserving finite volume method for solving the MCT governing equation is presented and verified for its second-order spatial accuracy. The fluxes are constructed using the generalized Lax-Friedrichs splitting scheme. A MCT-based method for inserting turbulent fluctuations into the fluid flow allows for the direct input of turbulent kinetic energy into the flow is also presented. When comparing MCT simulation data with the experiments of Kuntz et al \cite{kuntz1985experimental}, the MCT solver is shown to reproduce the surface pressure and velocity profile after the presence of the shock. The required cell number for simulation is compared with a DNS study in a similar case. The comparison shows MCT can provide meaningful results with the smallest cell size ($\Delta y^+$) ten times larger than the one used in the classical DNS. This comparison validates MCT as a computation-friendly alternative theory for compressible turbulence. 

A new statistical averaging procedure relying on the multiscale nature of MCT is also introduced and used to analyze energy cascade at the length scale of eddies. Through the newly introduced variable of subscale eddy rotation, the evolution of subscale eddy kinetic energy can be carefully monitored in a compressible turbulent flow. The results show that the fluctuations in the eddies' rotational energy correspond well to the fluctuations in the translational and internal energy, indicating a transfer of the subscale energy across the fluctuating components of the flow. These figures give a visual representation of the contribution of individual eddies to the overall dynamics of the turbulence, as well as its structure. A closer look at more complex compressible flows can assess features such as the effects of compressibility on subscale energy transfer. These simulations are left for future work.

\section*{Acknowledgement}
This material is based upon work supported by the Air Force Office of Scientific Research under award number FA9550-17-1-0154. 

\bibliography{bibtex_database}

\end{document}